\documentclass[runningheads]{llncs}

\usepackage[mobile]{eccv}

\usepackage{eccvabbrv}

\usepackage{graphicx}
\usepackage{booktabs}
\usepackage{wrapfig}
\usepackage{makecell}
\usepackage{booktabs} 

\usepackage[accsupp]{axessibility}  

\usepackage{hyperref}

\usepackage{orcidlink}

\begin{document}

\title{Coherent Audio-Visual Editing via Conditional Audio Generation Following Video Edits} 

\titlerunning{Coherent Audio-Visual Editing}

\author{Masato Ishii\inst{1} \and
Akio Hayakawa\inst{1} \and
Takashi Shibuya\inst{1} \and
Yuki Mitsufuji\inst{1,2}}

\authorrunning{M.~Ishii et al.}

\institute{Sony AI, Japan \and
Sony Group Corporation, Japan \\
\email{\{masato.a.ishii, akio.hayakawa, takashi.tak.shibuya, yuhki.mitsufuji\}@sony.com} }

\maketitle

\begin{abstract}

We introduce a novel pipeline for joint audio-visual editing that enhances the coherence between edited video and its accompanying audio. Our approach first applies state-of-the-art video editing techniques to produce the target video, then performs audio editing to align with the visual changes. To achieve this, we present a new video-to-audio generation model that conditions on the source audio, target video, and a text prompt. We extend the model architecture to incorporate conditional audio input and propose a data augmentation strategy that improves training efficiency. Furthermore, our model dynamically adjusts the influence of the source audio based on the complexity of the edits, preserving the original audio structure where possible. Experimental results demonstrate that our method outperforms existing approaches in maintaining audio-visual alignment and content integrity.

\end{abstract}    
\section{Introduction}
\label{sec:intro}

Recent advancements in video editing have enabled sophisticated manipulation of visual content~\cite{jiang2025vace,kara2024rave}. However, these techniques typically address only the visual modality, overlooking the crucial role of audio in most real-world videos. As a result, there is a growing demand for editing solutions that maintain strong audio-visual coherence. The emergence of audio-visual generation models, such as Veo 3~\cite{veo3} and Sora 2~\cite{sora2}, underscores the increasing importance of synchronized audio and video in content creation.

Existing approaches to joint audio-video editing face significant limitations. Editing audio and video modalities independently often results in poor alignment, as changes in one stream may not correspond to changes in the other. Alternatively, applying video-to-audio models to edited videos can disrupt the integrity of the original audio. To address these issues, joint editing methods have been studied~\cite{liang2024language,lin2025zero}, but they are restricted to low frame-rate videos (e.g., 4 fps~\cite{lin2025zero}), which is impractical for most applications. In contrast, modern video editing methods can process much higher frame rates (e.g., 16 fps~\cite{jiang2025vace}), but lack integrated audio editing capabilities.

\begin{figure}[t]
    \centering
    \includegraphics[width=0.9\linewidth, trim=0pt 3.5cm 0pt 0pt, clip]{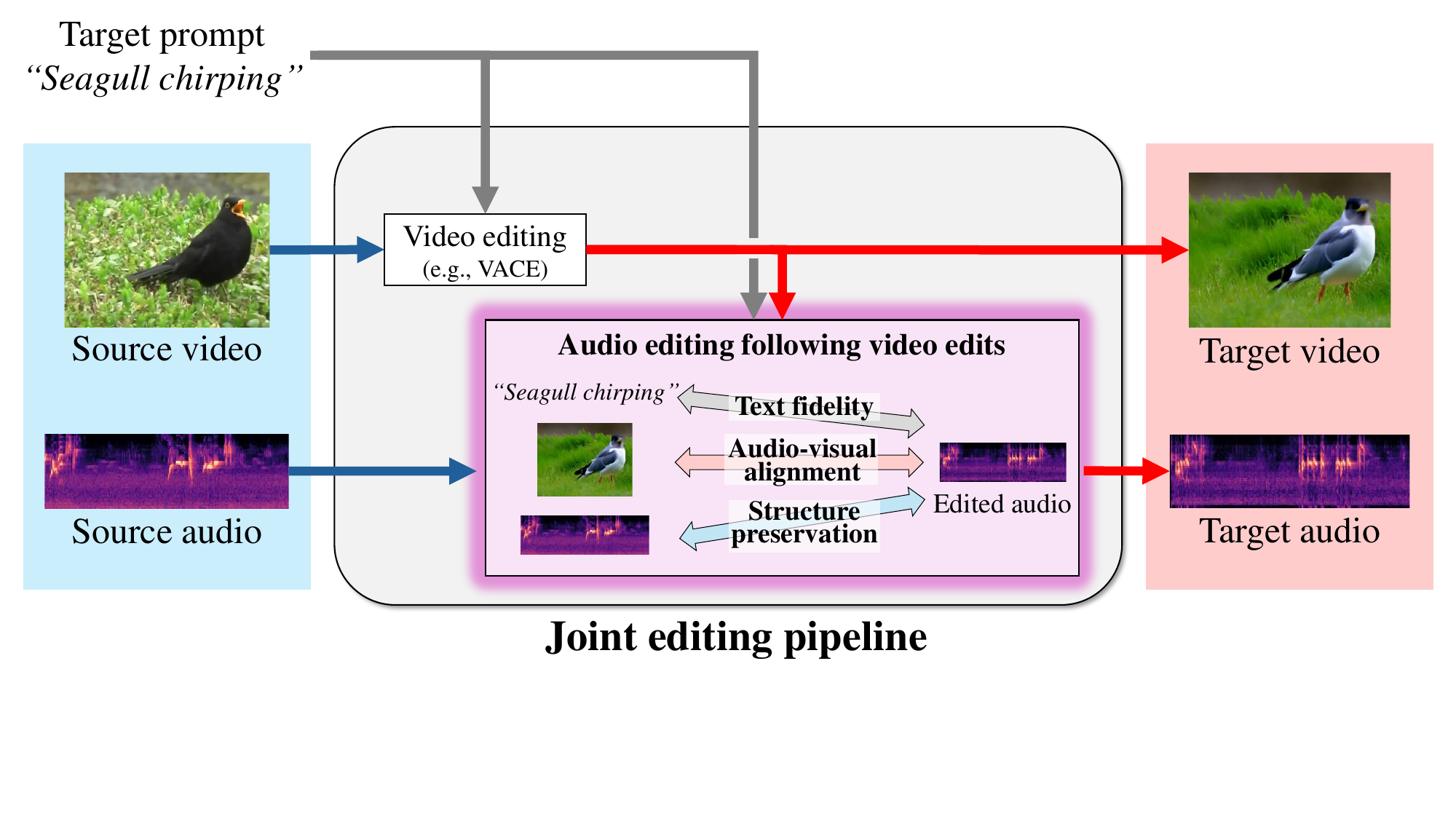}
    \caption{An overview of the proposed pipeline.}
    \label{fig:overview}
\end{figure}

To address these challenges, we decouple joint audio-video editing into two sequential sub-tasks as shown in Fig. \ref{fig:overview}. We first apply an off-the-shelf video editing method to transform the source video into the target video, leveraging recent progress in high frame-rate video editing without modification. We then introduce a new task setting, {\bf audio editing following video edits}, where the goal is to edit the source audio to align with the edited video while preserving the original audio structure as much as possible (e.g., the timing and number of sound events, and the continuity of background ambience), guided by a text prompt. This task is particularly challenging because the model must jointly satisfy text fidelity, audio-visual alignment to the edited visual events, and audio structure preservation.

For this new task, we propose a conditional video-to-audio editing model built by minimally extending a state-of-the-art video-to-audio generation model~\cite{cheng2025mmaudio}. Specifically, we modify the model to accept the source audio as an additional condition alongside the target video and text, enabling it to preserve the original audio structure. We further introduce a data augmentation strategy for this new conditional input to improve training efficiency, and we also adopt an adaptive conditioning mechanism to adjust the influence of the source audio based on the complexity of the edits, maximizing audio structure preservation. With these changes, our model can dynamically balance text fidelity, audio-visual alignment, and audio structure preservation. Experimental results demonstrate that our pipeline achieves superior performance compared to existing methods, particularly in maintaining audio-visual alignment and content integrity.

\section{Preliminaries and related work}

\subsection{Flow matching}
\label{sec:flow-matching}

Flow matching~\cite{lipman2023flow} is a framework for generative modeling that has been widely adopted across various domains due to its strong performance and high scalability. It employs a strategy essentially similar to diffusion models~\cite{ho2020denoising} but offers greater flexibility in model design~\cite{gao2025diffusion}. In this framework, the model learns a time-dependent velocity field. Let $\mathbf{x}_0$ and $\mathbf{x}_1$ denote samples from the prior and target distributions, respectively, and define $\mathbf{x}_t = t\mathbf{x}_0 + (1-t)\mathbf{x}_1$ as the linear interpolation between them at time $t \in [0,1]$. The model $u_\theta (\mathbf{x}, t)$, parameterized by $\theta$, predicts the velocity at each time $t$.

The model is trained by minimizing the conditional flow-matching objective:
\begin{align}
\label{eq:flow-matching}
    \min_\theta \mathbb{E}_{p(\mathbf{x}_0),p(\mathbf{x}_1),\mathcal{U}(t)} \| u_\theta (\mathbf{x}_t, t) - u(\mathbf{x}_t | \mathbf{x}_0, \mathbf{x}_1) \|^2,
\end{align}
where $u(\mathbf{x}_t | \mathbf{x}_0, \mathbf{x}_1)$ is the conditional velocity. In the current setting, where $\mathbf{x}_t$ is defined as a linear interpolation between $\mathbf{x}_0$ and $\mathbf{x}_1$, the conditional velocity simplifies to $u(\mathbf{x}_t | \mathbf{x}_0, \mathbf{x}_1) = \mathbf{x}_1 - \mathbf{x}_0$.

Given a trained flow-matching model, we can define the following ordinary differential equation (ODE):
\begin{align}
    \frac{\mathrm{d} \mathbf{x}_t}{\mathrm{d} t} = u_\theta (\mathbf{x}_t, t).
\end{align}
By solving this ODE from $t=0$ (with initial sample $\mathbf{x}_0 \sim p(\mathbf{x}_0)$) to $t=1$, we can transform a sample from the prior distribution into a sample from the target distribution $p(\mathbf{x}_1)$. By choosing a tractable distribution (e.g., Gaussian) for $p(\mathbf{x}_0)$ and the target data distribution for $p(\mathbf{x}_1)$, this model can serve as a generative model.

When modeling conditional generation, the model is extended to accept a conditional input $\mathbf{C}$, resulting in $u_\theta (\mathbf{x}_t, t, \mathbf{C})$, and the training objective in Eq. (\ref{eq:flow-matching}) is modified to sample from the joint distribution $p(\mathbf{x}_1, \mathbf{C})$ instead of $p(\mathbf{x}_1)$. During inference, classifier-free guidance~\cite{ho2021classifier} can be applied to the velocity prediction to improve fidelity to the given condition:
\begin{align}
    \tilde{u}_\theta (\mathbf{x}_t, t, \mathbf{C}) =\ & u_\theta (\mathbf{x}_t, t, \varnothing) + w (u_\theta (\mathbf{x}_t, t, \mathbf{C}) - u_\theta (\mathbf{x}_t, t, \varnothing)),
\end{align}
where $\varnothing$ denotes the null condition, and $w$ is a coefficient to control the strength of the guidance. To enable the model to handle the null condition, the conditional input is randomly dropped with a small probability (e.g., 0.1) during training.

\subsection{Single-modal data editing}

Editing refers to the process of creating target data by modifying specific content in given source data, as intended by the user, while preserving unrelated elements. This section briefly reviews editing methods based on flow matching or diffusion models in both audio and video domains.



\paragraph{Audio Editing.}  
Audio editing methods can be broadly categorized into training-based and training-free approaches. In the training-based approach~\cite{wang2023audit,lan2025guiding}, models are trained as conditional generators that take source audio as input and produce edited outputs. This approach is conceptually straightforward, but it requires large-scale datasets of paired source and target audio samples, which is challenging in practice. In contrast, training-free approaches~\cite{manor2024zero,xu2024prompt,jia2025audioeditor} leverage pre-trained text-to-audio generation models~\cite{liu2023audioldm,liu2024audioldm,ghosal2023text} to generate the desired audio edits. Although they eliminate the need for paired training datasets, they require techniques to inject source audio information into the generation process, which typically incurs significantly higher computational costs during generation. Our method is also training-based one, but we mitigate the dataset issue by adopting acoustic feature extraction as described in Section \ref{sec:feature}.

\paragraph{Video Editing.}  
Training video generation models from scratch is considerably more resource-intensive than audio models. As a result, most video editing methods rely on pre-trained generative models. There are two main types of pre-trained models used: text-to-image and text-to-video models.

Early work leveraged text-to-image models such as Stable Diffusion~\cite{rombach2022high} due to their accessibility and ease of use. Since these models are only capable of generating single frames, specialized techniques~\cite{cong2024flatten,li2024vidtome,cohen2024slicedit,geyer2024tokenflow,kara2024rave} and fine-tuning strategies~\cite{wu2023tune,liu2024video} have been developed to improve temporal consistency across video frames.
Recently, the emergence of advanced text-to-video generation models~\cite{blattmann2023stable,yang2025cogvideox,wan2025wan} has enabled more effective video editing. Studies utilizing these models~\cite{fan2024videoshop,shi2024bivdiff,zhang2025v2edit,jiang2025vace} demonstrate excellent temporal consistency.









\subsection{Joint audio-video editing}

Joint audio-video editing presents unique challenges compared to single-modal editing. Editing audio and video independently often results in poor alignment between the generated modalities, degrading the overall audio-visual coherence in the edited output. Therefore, ensuring strong alignment between audio and video is a central focus in this task, and it has been addressed in only a few recent studies~\cite{liang2024language,lin2025zero}.

Liang \etal~\cite{liang2024language} introduced a lightweight module that embeds the source audio-video pair into a specialized text token. This token is then input into pre-trained text-to-image and text-to-audio models to generate the corresponding video frames and audio. In contrast, Lin \etal~\cite{lin2025zero} extended score distillation sampling to the cross-modal setting by jointly leveraging pre-trained text-to-image and text-to-audio models. They further incorporated a contrastive loss on relevant and irrelevant regions within the audio-video pair to improve consistency between audio and video edits.

However, these approaches share two main limitations. First, they depend on text-to-image models for video editing, which restricts the ability to maintain temporal consistency across video frames. Second, they lack mechanisms to synchronize the temporal alignment between edited audio and video, making precise audio-visual alignment difficult. Thus, their application is essentially limited to low frame-rate videos (e.g., 1 fps~\cite{liang2024language} or 4 fps~\cite{lin2025zero}).

We address these limitations with a sequential approach to joint editing. First, we apply a state-of-the-art video editing technique to generate the target video, ensuring high temporal consistency. Next, we use our novel video-to-audio model, which incorporates the source audio as an additional conditional input, to conduct audio editing following the video edits, thereby achieving strong audio-visual alignment in much higher frame-rate videos (20 fps in the experiments).

\section{Proposed method}

In this section, we first show an overview of our method and briefly explain how it works. Then, we describe the details of the newly introduced mechanisms designed for the video-guided audio editing task.

\subsection{Overview}

Our goal is to build a model that generates edited audio aligned with the target edited video while preserving the acoustic structure of the source audio. To achieve this, we first extract hierarchical acoustic features from the source audio to capture its structural information at multiple levels of detail (Section \ref{sec:feature}). These features, together with conditional features from the target video and a target prompt, are fed into the model (Section \ref{sec:architecture}), which is trained to generate the edited audio using a flow matching framework. To improve training, we introduce a new data-augmentation strategy for the acoustic features called detail-temporal masking (Section \ref{sec:masking}). During generation, the model automatically selects the appropriate level of detail for the acoustic features based on the estimated editability score, which is determined by measuring the semantic similarity between the source audio and the target video (Section \ref{sec:adaptive_conditioning}).

\subsection{Hierarchical acoustic feature extraction}
\label{sec:feature}

\paragraph{Motivation.}
To preserve the acoustic structure of the source audio, we need to extract this information from it. In our setting, the feature representation should satisfy the following three properties:
\begin{itemize}
    \item The feature effectively represents the acoustic structure information.
    \item The feature is robust to semantic changes during editing, ensuring it remains informative for generating the edited audio.
    \item The feature can be extracted at multiple levels of detail, enabling control over the amount of preserved structural information.
\end{itemize}
A common choice to extract audio features would be using a pretrained model. However, such a model generally does not satisfy the second property, as they are typically trained to be sensitive to differences in audio semantics (e.g., CLAP~\cite{elizalde2023clap}). As a result, when used as “structure” features, they tend to overly preserve the original semantic content of the source audio (e.g., specific sound identities), which hinders semantic edits and can cause the editing to fail. In addition, extracting semantic features at a specified level of detail is challenging, which conflicts with the third property. Consequently, we design simpler statistical features of audio signals that satisfy all three properties. Statistical features, such as loudness, are less sensitive to semantic changes in audio content, making them more reliable for preserving structural information~\cite{garcia2025sketch2sound}.

\paragraph{Implementation.}
Inspired by Sketch2Sound~\cite{garcia2025sketch2sound}, our acoustic feature is based on the frame-wise loudness of the audio signal processed with a median filter for temporal smoothing. We follow a common definition of loudness, which is an A-weighted sum across the frequency bins in a magnitude spectrogram~\cite{morrison2024fine}. To extract acoustic features at various levels of detail, we recursively split the frequency bins of the magnitude spectrogram into two sub-ranges (e.g., low and high frequencies) and compute the loudness for each sub-range at every level. Let $\mathbf{s} \in \mathbb{R}^{F \times T_a}$ denote the spectrogram of the audio, where $F$ is the number of frequency bins and $T_a$ is the number of time-frames. At level $l$, $\mathbf{s}$ is divided into $\{\mathbf{s}_i^{(l)} \in \mathbb{R}^{F/2^l \times T_a} \}_{i=0,...,2^l-1}$, and the acoustic features are calculated as follows:
\begin{align}
    &\mathbf{a}_l = \mathrm{concat} \left( \left\{ A ( \mathbf{s}_i^{(l)} ) \right\}_{i=0}^{2^l-1} \right),
\end{align}
where $A$ is a function that computes the A-weighted sum, and $\mathrm{concat}$ denotes concatenation along the frequency axis.

\begin{wrapfigure}{r}{0.50\linewidth}
    \centering
    \includegraphics[width=0.99\linewidth, trim=0pt 6cm 0pt 0pt, clip]{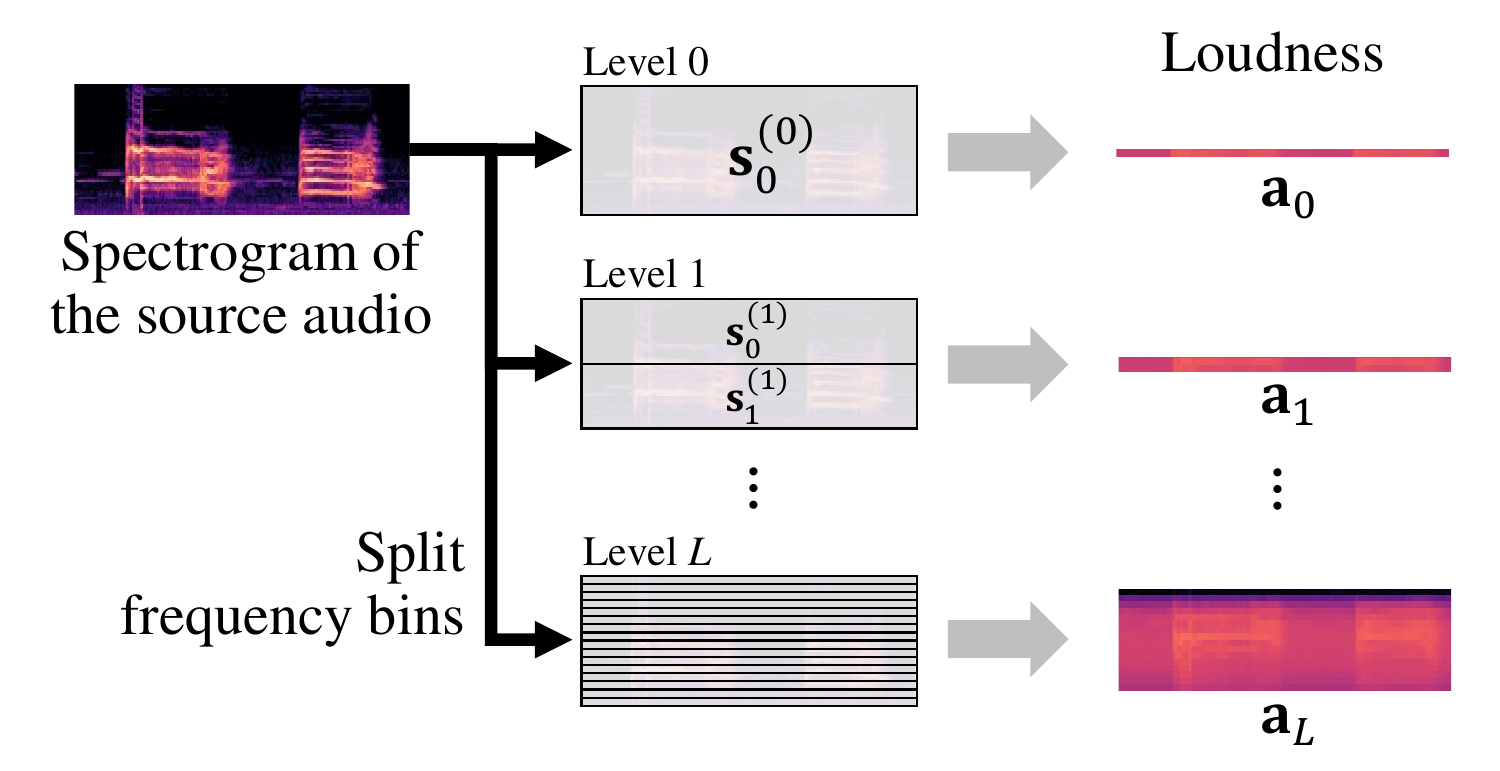}
    \caption{Hierarchical acoustic features.}
    \label{fig:feature}    
\end{wrapfigure}

As shown in Fig. \ref{fig:feature}, this process yields a hierarchy of acoustic features $\{ \mathbf{a}_l \}$, from coarse to fine detail, allowing flexible control over structural information preservation. The level of detail can be controlled by masking out the features corresponding to finer details. Specifically, the acoustic features at the specified level of detail are given by:
\begin{align}
    &\mathbf{a} = \mathrm{concat} \left( \{ \mathbf{a}_l \odot \mathbf{m}_l, \mathbf{m}_l \}_{l=0}^L \right),
\end{align}
where $L$ is the maximum level, and $\mathbf{m}_l \in \mathbb{R}^{2^l \times T_a}$ denote a mask at level $l$, whose elements are all zero if $l$ is larger than the specified level, and all one otherwise.
In the remainder of this paper, we refer to $\mathbf{a}$ as the acoustic features.

\subsection{Model architecture}
\label{sec:architecture}

\begin{figure*}[t]
    \centering
    \includegraphics[width=0.85\linewidth]{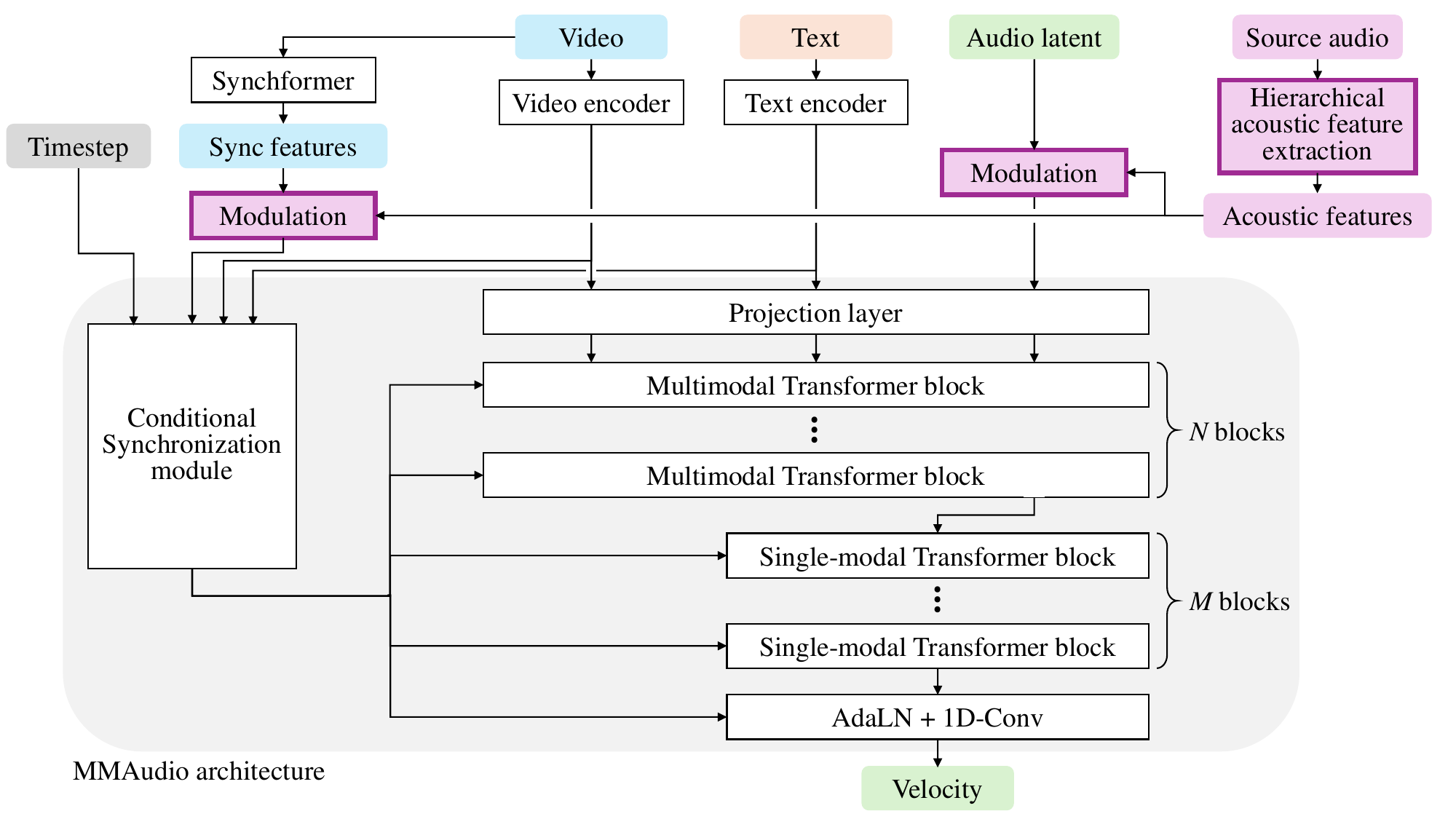}
    \caption{The architecture of our model. The major modifications from MMAudio are shown in purple.}
    \label{fig:model}
\end{figure*}

Our model shown in Fig. \ref{fig:model} is based on the MMAudio architecture~\cite{cheng2025mmaudio}, with two key modifications to effectively handle information from the acoustic features.

\paragraph{Base architecture.}
The backbone of the base model is a multi-modal Transformer that processes noisy audio latent features, as well as conditional text and video features, to predict the velocity for audio generation through the flow matching framework. 
It comprises a sequence of MM-DiT blocks~\cite{esser2024scaling} followed by a sequence of audio-only single-modal DiT blocks. Additionally, the base model includes a conditional synchronization module to enhance audio-visual synchrony. This module extracts high frame-rate features from the input video using Synchformer~\cite{iashin2024synchformer} and combines them with temporally-global conditional features extracted from text and video. The combined features are fed into the DiT blocks through adaptive layer normalization~\cite{perez2018film}.

\paragraph{Modifications.}
To incorporate the acoustic features, we introduce two modifications to the base model:
\begin{itemize}
    \item \textbf{Modulation of audio latents:} We add the processed acoustic features to the audio latents. Before the addition,  we adjust the temporal length of the acoustic features by linear interpolation and process them with a trainable linear layer to match the dimensionality of the audio latents. This approach, which uses input addition or concatenation, is a common method for integrating conditional structural data into flow matching models and has been widely used in prior studies~\cite{wang2024frieren,zhu2024flowie}.
    \item \textbf{Modulation of Synchformer features:} We also modulate the Synchformer features in a frame-wise manner using the acoustic features, which are processed with temporal length regulation and a feed-forward module (as in the DiT block). This temporally-local modulation is important because the Synchformer features strongly influence the generated audio in terms of temporal dynamics via adaptive layer normalization. To properly reflect the conditional acoustic information in the generation process, it is essential to modulate the Synchformer features before they are used in the base model.
\end{itemize}

These modules are initialized so that the modulation simply results in an identical function. During training, they are fixed in the first half of the training iterations and are set trainable there after. We adopt this two-stage training strategy to prevent the model from too much focusing on the acoustic features rather than the text prompt and the target video to generate the target audio.

\subsection{Training with detail-temporal masking}
\label{sec:masking}

\begin{figure}[t]
  \centering
  \begin{minipage}[b]{0.48\linewidth}
    \centering
    \includegraphics[width=0.9\linewidth]{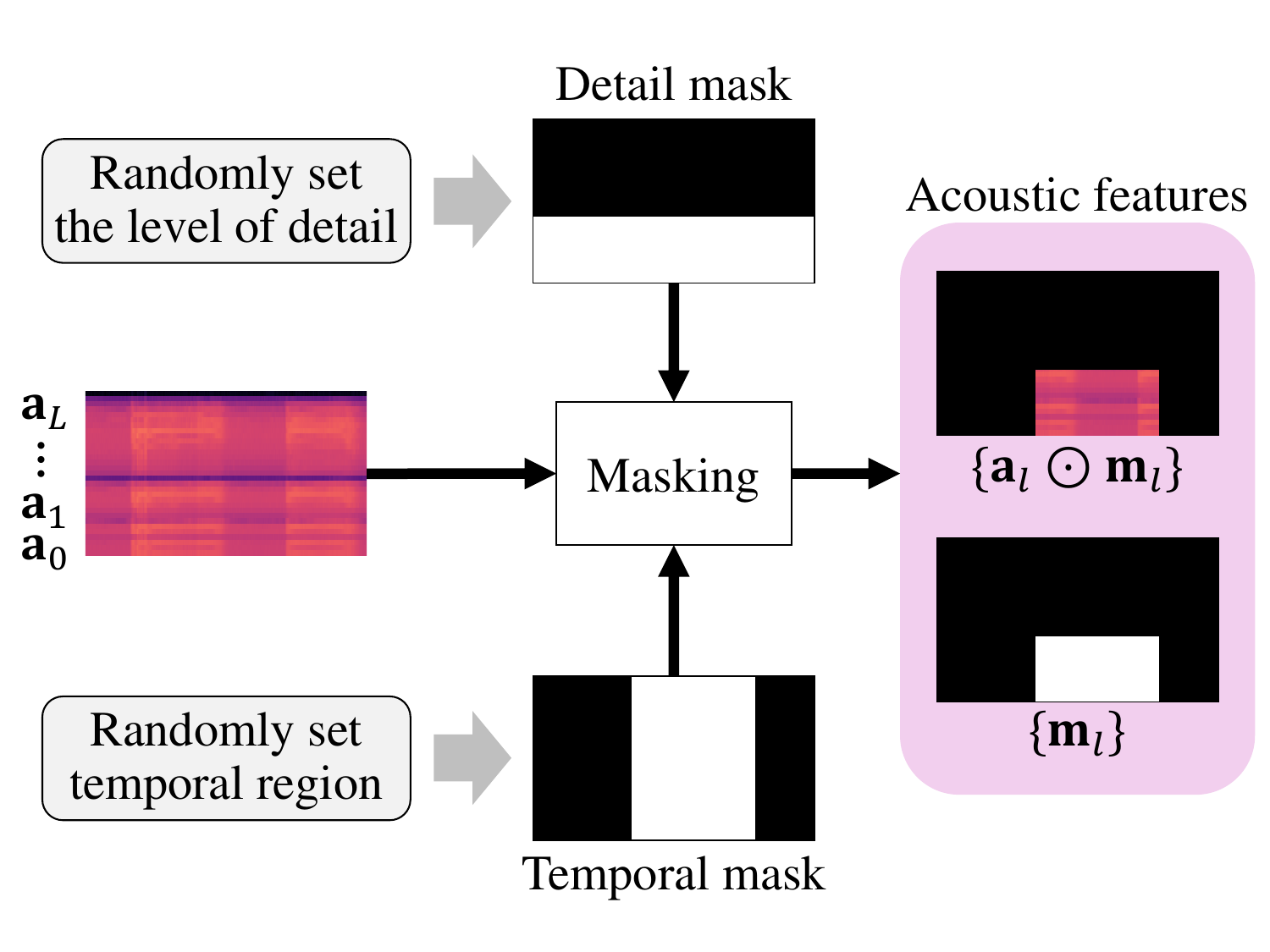}
    \caption{Detail-temporal masking of the acoustic features during training.}
    \label{fig:masking}
  \end{minipage}\hfill
  \begin{minipage}[b]{0.48\linewidth}
    \centering
    \includegraphics[width=0.9\linewidth]{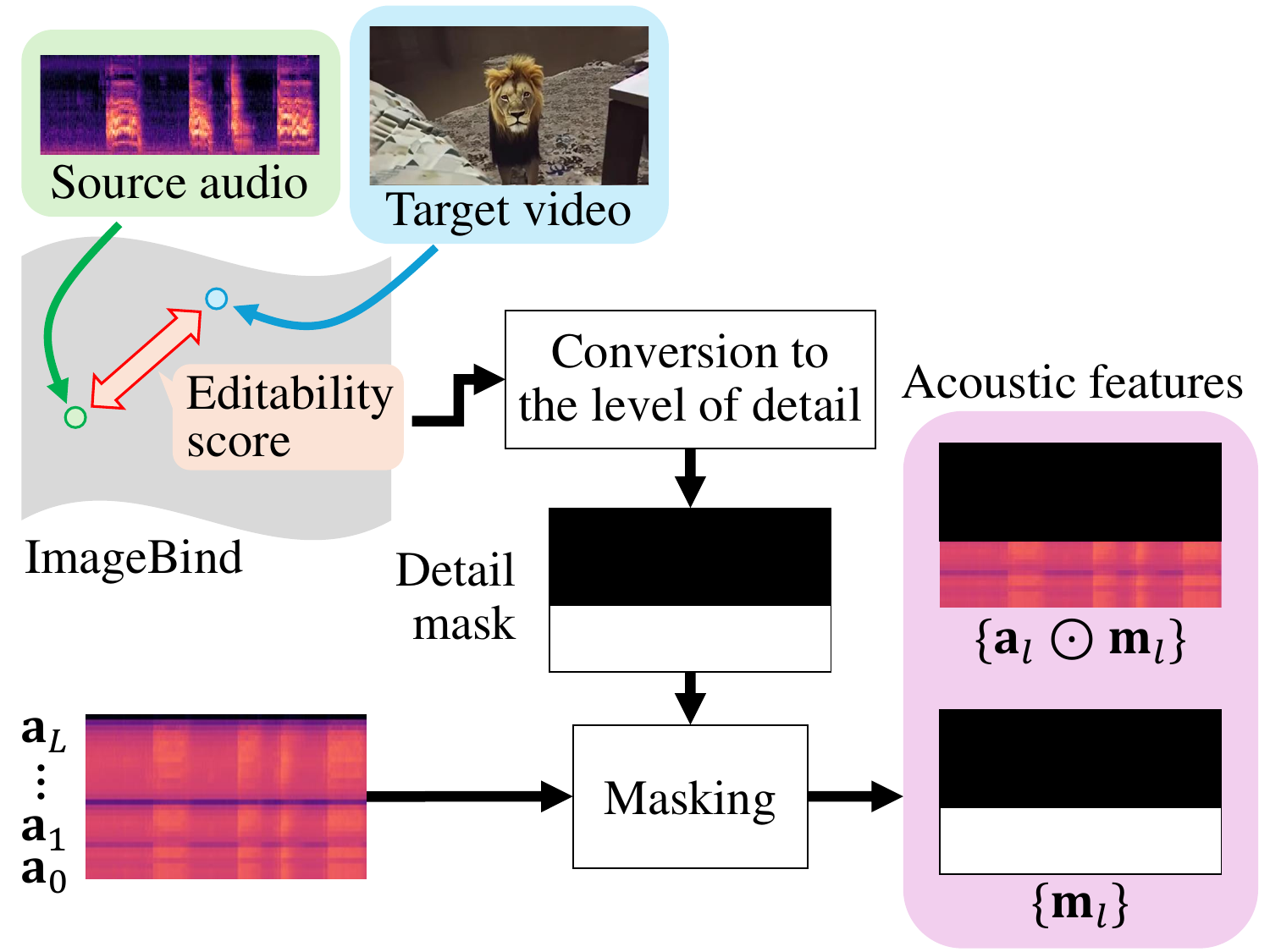}
    \caption{The adaptive conditioning based on the editability score.}
    \label{fig:conditioning}
  \end{minipage}
\end{figure}

We train our model through flow matching described in Section \ref{sec:flow-matching}. 
Ideally, training would require sets of source audio, text prompt, target video, and target audio for each sample, but collecting such paired data with accompanying text and video at scale is challenging. Instead, we use standard text-audio-video data as a triplet of text prompt, target audio, and target video. To mimic the acoustic features of the source audio, we extract them from the target audio. Since the acoustic features are robust to semantic differences, as explained in Section 3.2, we expect the model to learn effectively from such training data.

We jointly apply two kinds of masking to augment the acoustic features as shown in Fig. \ref{fig:masking}: detail masking and temporal masking. These two masking processes are introduced for separate purposes as described below. 
\begin{itemize}
    \item \textbf{Detail masking.} For each training sample, we randomly determine the level of detail of the acoustic features. Since this corresponds to randomly choosing a proper mask for the acoustic features as described in Section \ref{sec:feature}, we refer to this augmentation as detail masking. This masking enables the model to work with various levels of details, which means that we can control how much the original acoustic structure in the source audio should be preserved in the target audio at the inference time. We also include a choice to totally mask out the entire acoustic features in the setting, enabling classifier-free guidance for more accurate structure preservation.
    \item \textbf{Temporal masking.} We also apply random masking along with the temporal dimension to prevent the model from excessively relying on the acoustic features to generate the target audio. Since the acoustic features contain temporally rich information of the audio, especially when the level of detail is high, the model often focus more on aligning with the acoustic features rather than text and video. This leads to degrading text fidelity and audio-visual alignment, which is not desirable for the audio editing. To alleviate this issue, we mask out the acoustic features at randomly selected time frames during training. It enforces the model to jointly reflect both the acoustic features and the other conditional features to generate the target audio.
\end{itemize}

\subsection{Generation with adaptive conditioning}
\label{sec:adaptive_conditioning}

Once the model is trained, we can generate the target audio by solving the ODE as described in Section \ref{sec:flow-matching}. 
Following the prior works on multi-conditional generation~\cite{brooks2023instructpix2pix,chen2024gentron,kushwaha2025vintage}, we adopt classifier-free guidance with multiple guidance terms as shown below:
\begin{align}
    \tilde{u}_\theta (\mathbf{x}_t, t, \mathbf{C}, \mathbf{a}) = u_\theta (\mathbf{x}_t, t, \varnothing, \varnothing) &+ w_1 ( u_\theta (\mathbf{x}_t, t, \mathbf{C}, \varnothing) - u_\theta (\mathbf{x}_t, t, \varnothing, \varnothing)) \nonumber \\
    & + w_2 (u_\theta (\mathbf{x}_t, t, \mathbf{C}, \mathbf{a}) - u_\theta (\mathbf{x}_t, t, \mathbf{C}, \varnothing)),
\end{align}
where $\mathbf{a}$ and $\mathbf{C}$ represent the acoustic features extracted from the source audio and a pair of text and video conditions, respectively, and $w_1$ and $w_2$ are coefficients to control the strength of two guidance terms. The first guidance term enhances the fidelity to the text and video conditioning, while the second term improves the structure preservation through referring to the acoustic features. In the experiments, we used $w_1=7.5$ and $w_2=0.5w_1$.

During generation, the model adaptively selects the appropriate level of detail for the acoustic features depending on the estimated difficulty of the audio editing task. This selection is performed deterministically based on an editability score, as described below. We term this process adaptive conditioning.

\paragraph{Motivation of adaptive conditioning.} The optimal level of detail depends on how difficult the audio editing task is. For easy edits, where the source audio is already similar to the target (e.g., bird chirping $\rightarrow$ seagull chirping), preserving detailed acoustic features helps maintain structural fidelity without sacrificing alignment with the target prompt and video. In contrast, for difficult edits, where significant changes are needed (e.g., cat meowing $\rightarrow$ lion roaring), retaining too much detail from the source is less effective, much of the original acoustic structure will be altered or lost. Therefore, our method adaptively adjusts the level of acoustic detail based on the estimated editing difficulty for each sample, aiming to balance structural preservation and target fidelity.

\paragraph{Implementation of adaptive conditioning.} 
To implement adaptive conditioning, we estimate an editability score that represents how easy the audio editing is expected to be. Specifically, we compute the audio-visual semantic similarity between the source audio and the target video. We use ImageBind~\cite{girdhar2023imagebind} to embed the audio and video clips in a common feature space and then calculate the cosine similarity between these embeddings. Since ImageBind can only handle relatively short clips (e.g., two seconds), we use a sliding window approach with a window size of two seconds without overlapping, repeating this process and averaging the similarities across all time frames.

Given the editability score $s$, we determine the level of detail by quantizing the normalized score as shown below.
\begin{align}
\label{eq:l}
    l = \mathrm{round} \left( l_\mathrm{max} \cdot \mathrm{clamp} \left( \frac{s-s_\mathrm{min}}{s_\mathrm{max}-s_\mathrm{min}} \right) \right),
\end{align}
where $s_\mathrm{min}$ and $s_\mathrm{max}$ are the expected maximum and minimum scores, respectively, and $l_\mathrm{max}$ denote the maximum level of detail. For example, an easy edit ($s$ close to $s_\mathrm{max}$) results in a higher level of detail being preserved, while a difficult edit ($s$ close to $s_\mathrm{min}$) results in less detail being retained.

We control the structure preservation in our model by changing the value of $l_\mathrm{max}$. Allowing a larger level of detail leads to more accurate preservation of the original audio structure. On the other hand, we fix both $s_\mathrm{max}$ and $s_\mathrm{min}$. To determine these values, we empirically investigated the editability scores of genuine pairs of audio and video as well as those of random pairs in the VGGSound validation dataset~\cite{chen2020vggsound}, which can be considered as the maximum and minimum levels of the editability score for this dataset. Their average scores are $0.32$ and $0.02$, and these are used to set $s_\mathrm{max}$ and $s_\mathrm{min}$, respectively. 


\section{Experiments}

\subsection{Dataset and evaluation metrics}

\paragraph{Training datasets.} Following MMAudio~\cite{cheng2025mmaudio}, we jointly used several text-video-audio and text-audio datasets to train our model: VGGSound~\cite{chen2020vggsound}, AudioCaps~\cite{kim2019audiocaps}, Clotho~\cite{drossos2020clotho}, and WavCaps~\cite{mei2024wavcaps}. All audio clips were resampled to 16 kHz, and 8-second segments were cropped without overlap for each training sample.

\paragraph{Evaluation dataset.} We used the AvED-Bench dataset~\cite{lin2025zero}, which has particularly been constructed to evaluate audio-visual editing performance. The AvED-Bench dataset consists of 110 ten-second clips covering various types of natural audio-visual events. For each clip, a pair of the source and target prompts is given to specify the editing task. Audio and video were resampled to 16 kHz and 20 fps, respectively, and used as the source audio and video in our experiments.

\paragraph{Metrics.} We assess the generation quality from three different perspectives:
\begin{itemize}
    \item {\bf Audio-visual alignment (IB-AV)} assesses the alignment of the target audio to the target video. We used ImageBind to compute it as described in Section \ref{sec:adaptive_conditioning}. 
    \item {\bf Structure preservation (LPAPS)} assesses how much the acoustic structure of the source audio is preserved in the target audio. Following the prior studies~\cite{manor2024zero,lin2025zero}, we used LPAPS~\cite{iashin2021taming} to evaluate the structure preservation.
    \item {\bf Text fidelity (IB-TA)} assesses the fidelity of the target audio to the given target prompt using ImageBind. Since ImageBind can only handle two-second audio clips, we cropped five non-overlapping audio sub-clips from the target audio and extracted ImageBind features from each sub-clip. Then, we computed the cosine similarity between those features and the feature extracted from the prompt. To aggregate the multiple similarities, we used max pooling, because the audio event described in the prompt is often temporally local (e.g., dog barking) and thus could be missed in several sub-clips.
\end{itemize}

\subsection{Setup}

\begin{figure*}
    \centering
    \includegraphics[width=0.95\linewidth]{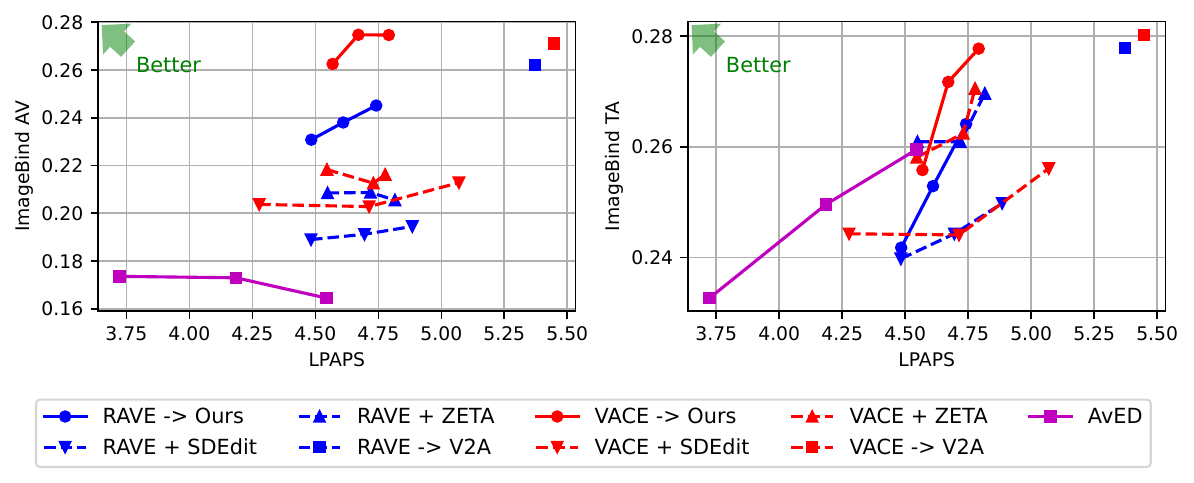}
    \caption{Experimental results on the AvED-Bench dataset. The horizontal axis represents structure preservation (lower is better). The vertical axis in the left plot indicates audio-visual alignment, while the right plot shows text fidelity (higher is better in both cases). For each method except $*\rightarrow$V2A, we varied the hyperparameters to obtain results with different degrees of structure preservation.}
    \label{fig:main_results}
\end{figure*}

We evaluated several methods representing three main approaches to audio-visual editing: independent, sequential, and joint.

\paragraph{Independent approach.} In this approach, audio and video are edited separately. For audio editing, we employed SDEdit~\cite{meng2022sdedit} and ZETA~\cite{manor2024zero}, both adapted to work with Stable Audio Open~\cite{evans2025stable}, a state-of-the-art audio generation model. For video editing, we used RAVE~\cite{kara2024rave}, which is based on Stable Diffusion 1.5~\cite{rombach2022high}, and VACE~\cite{jiang2025vace}, which utilizes the latest video generation model, Wan 2.1~\cite{wan2025wan}. Among the various editing tasks supported by VACE, we used the ``structure transfer" setting. To control the structure preservation in SDEdit and ZETA, we varied the intermediate timestep at which we inject the source audio information.

\paragraph{Sequential approach.} In this approach, audio and video are edited in sequence. Our proposed method belongs to this category: we apply it to videos processed by RAVE and VACE (RAVE/VACE $\rightarrow$ Ours). 
We also evaluated our method with fully masked acoustic features, which completely remove source audio information from conditioning. This setting corresponds to applying a naive video-to-audio model to the edited videos (RAVE/VACE $\rightarrow$ V2A).

\paragraph{Joint approach.} In this approach, audio and video are edited jointly. We chose AvED~\cite{lin2025zero} as the state-of-the-art baseline. We controlled the structure preservation by varying the guidance scale used in the score distillation. 

\subsection{Main results}

\begin{figure}[t]
    \centering
    \begin{minipage}[b]{0.48\linewidth}
        \centering
        \includegraphics[width=0.99\linewidth]{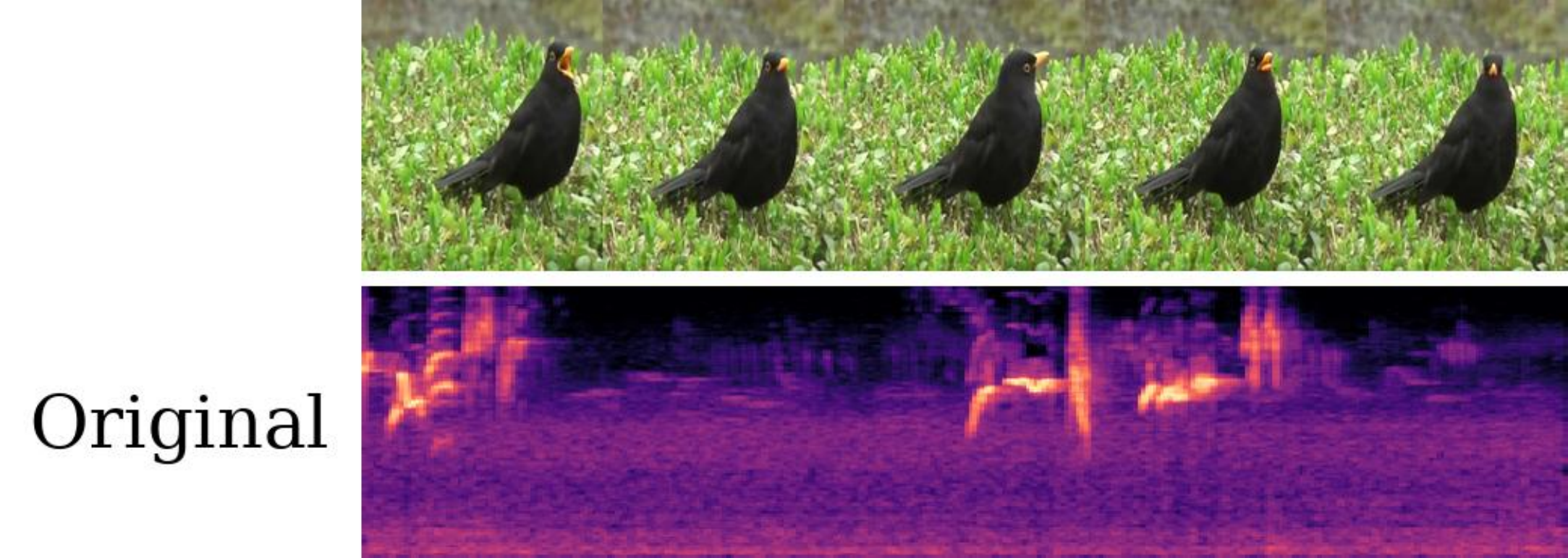}
        \subcaption{The source video and audio.}

        \vspace{\baselineskip}

        \includegraphics[width=0.99\linewidth]{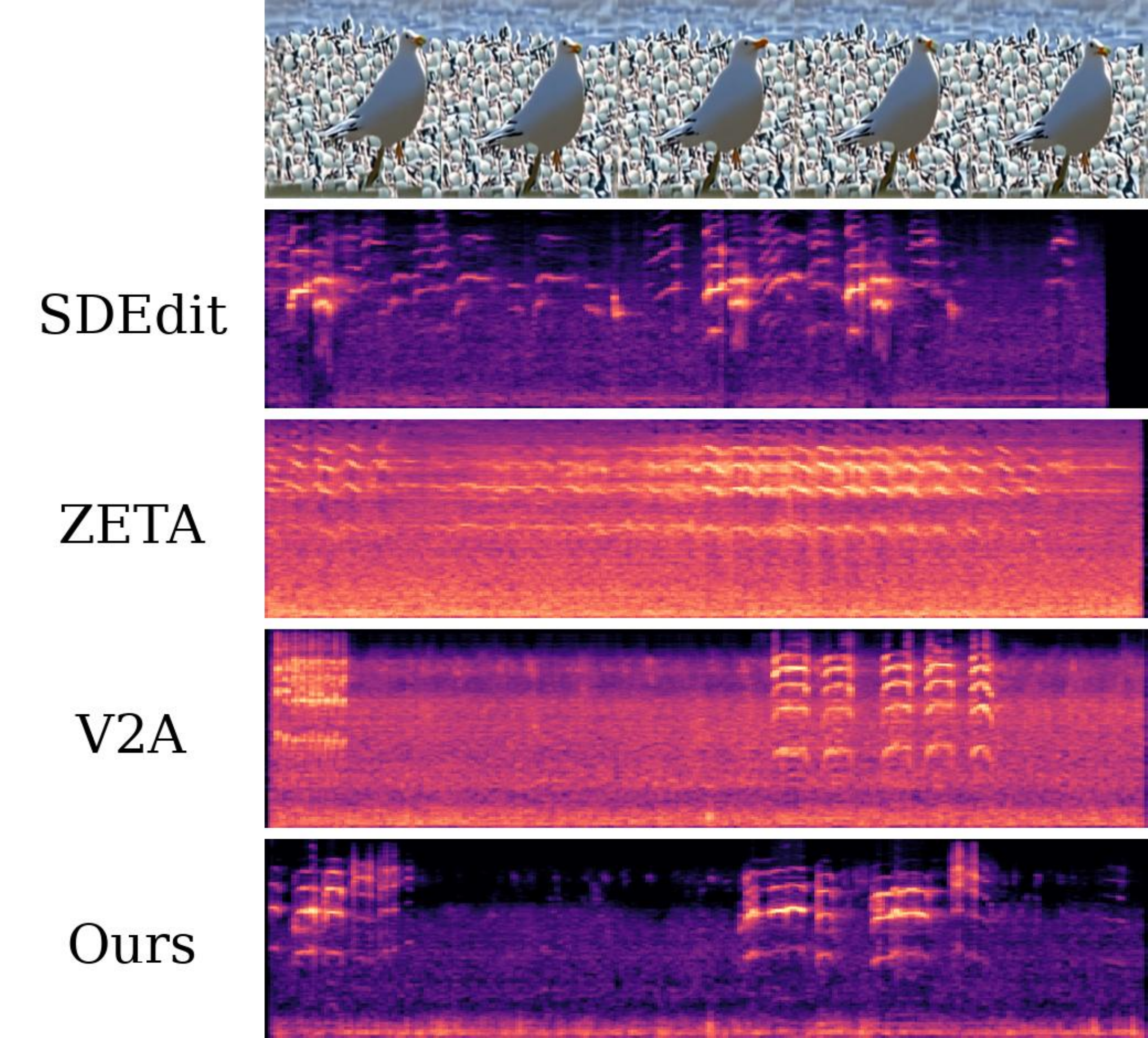}
        \subcaption{The target video edited by RAVE and the target audio generated by each method.}
    \end{minipage}\hfill
    \begin{minipage}[b]{0.48\linewidth}
        \centering
        \includegraphics[width=0.99\linewidth]{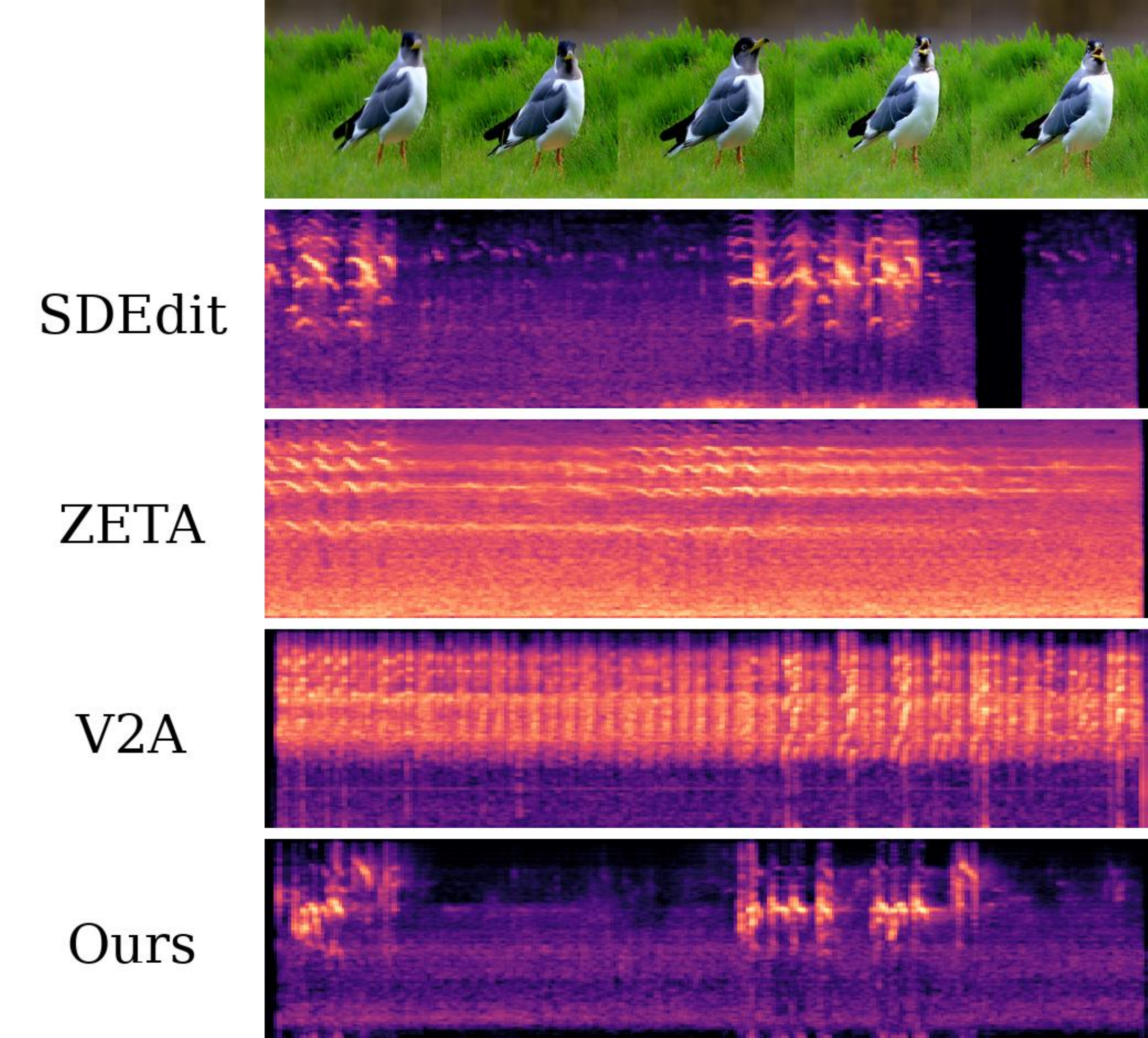}
        \subcaption{The target video edited by VACE and the target audio generated by each method.}

        \vspace{\baselineskip}

        \includegraphics[width=0.99\linewidth]{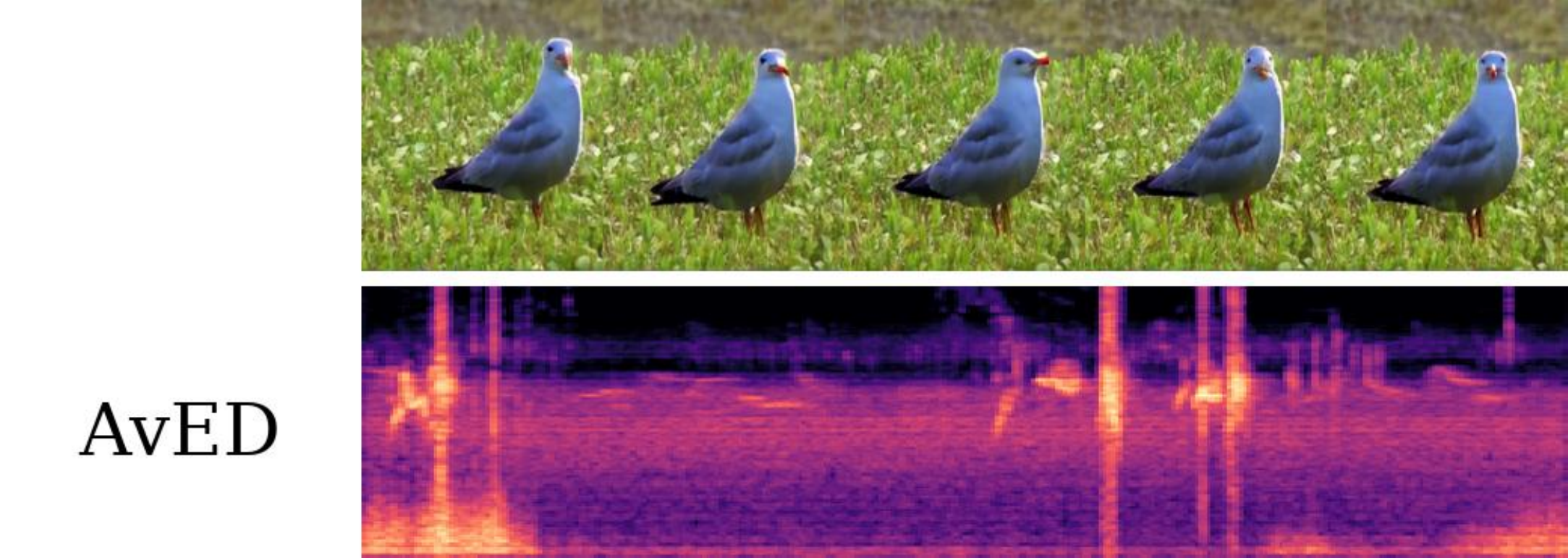}
        \subcaption{The target video edited by AvED.}
    \end{minipage}

    \caption{Examples of the generated audio and video.}
    \label{fig:example}
\end{figure}

Figure \ref{fig:main_results} summarizes the performance of all editing methods. In both plots, the horizontal axis indicates structural preservation (LPAPS). The left plot reports audio-visual alignment between the target audio and video (IB-AV), and the right plot reports target-audio text fidelity (IB-TA). Figure \ref{fig:example} shows qualitative examples. More examples are provided in the supplementary materials.

Here are several observations we obtained through this experiment:
\begin{itemize}
    \item {\bf Benefit of sequential approach.} The sequential editing methods achieved significantly better performance in audio-visual alignment as they leverage the latest video-to-audio architecture when generating the edited audio. In contrast, the independent editing methods achieved resulted in poor audio-visual alignment due to the lack of a mechanism for aligning edited audio and video. AvED also failed to achieve high audio-visual alignment due to the difficulty of jointly editing both audio and video.
    \item {\bf Advantage of our method.} Simply applying video-to-audio to the edited video results in poor structure preservation, because naive video-to-audio generation cannot take the source audio into account. On the other hand, our method effectively maintains the structure of the source audio while achieving high audio-visual alignment and text fidelity.
    \item {\bf Dependency on the video editing method.} Compared with VACE $\rightarrow$ ours, the advantage of RAVE $\rightarrow$ ours is marginal especially in terms of text fidelity. This is due to the relatively lower performance of RAVE in video editing. RAVE sometimes fails to generate a reasonable video that is well aligned with the given text prompt. In such cases, our model struggles to handle target videos and text prompts that semantically contradict each other, leading to degraded quality of the edited audio.
\end{itemize}

\subsection{Subjective test}

We also conducted a subjective test to evaluate the perceptual quality of the edited videos. In this test, each rater first watches and listens to the source video and then views each edited version. The rater is asked to rate the quality of each edited video on a scale from 1 (poor) to 5 (excellent) in terms of the following three aspects: audio-text fidelity, audio-visual alignment, and audio structure preservation. In total, eleven raters participated in the test, and each rater evaluated six different videos.

Table \ref{tab:user_study} shows the results of the subjective test. Our method achieved the highest scores across all metrics. Consistent with the main evaluation results (Fig. \ref{fig:main_results}), the proposed method and V2A achieved high audio-text fidelity and audio-visual alignment. In contrast, for audio structure preservation, unlike the main evaluation results, ZETA and AvED showed low scores. This suggests that, when the edited audio is experienced together with the video, perceived consistency with the video strongly influences subjective ratings.

\begin{table}[t]
    \centering
    \begin{tabular}{cccc}
        \toprule
         & \parbox{7em}{\centering audio-text\\ fidelity} & \parbox{7em}{\centering audio-visual\\ alignment} & \parbox{7em}{\centering audio structure\\ perservation} \\
         \midrule
        VACE $\rightarrow$ ZETA & 3.0 {\scriptsize($\pm$ 0.5)} & 2.6 {\scriptsize($\pm$ 0.7)} & 2.5 {\scriptsize($\pm$ 0.7)}  \\
        VACE $\rightarrow$ V2A & 3.2 {\scriptsize($\pm$ 0.6)} & 3.3 {\scriptsize($\pm$ 0.6)} & 3.2 {\scriptsize($\pm$ 0.5)} \\
        AvED & 2.7 {\scriptsize($\pm$ 0.4)} & 2.5 {\scriptsize($\pm$ 0.6)} & 3.1 {\scriptsize($\pm$ 0.4)} \\
        VACE $\rightarrow$ Ours & {\bf 3.7} {\scriptsize($\pm$ 0.5)} & {\bf 3.8} {\scriptsize($\pm$ 0.3)} & {\bf 3.9} {\scriptsize($\pm$ 0.2)} \\
        \bottomrule
    \end{tabular}
    \caption{Subjective evaluation results (mean score $\pm$ 95\% confidence interval). Bold indicates the best score for each metric.}
    \label{tab:user_study}
\end{table}

\subsection{Ablation study}

\begin{figure}[t]
    \centering
    \includegraphics[width=0.95\linewidth]{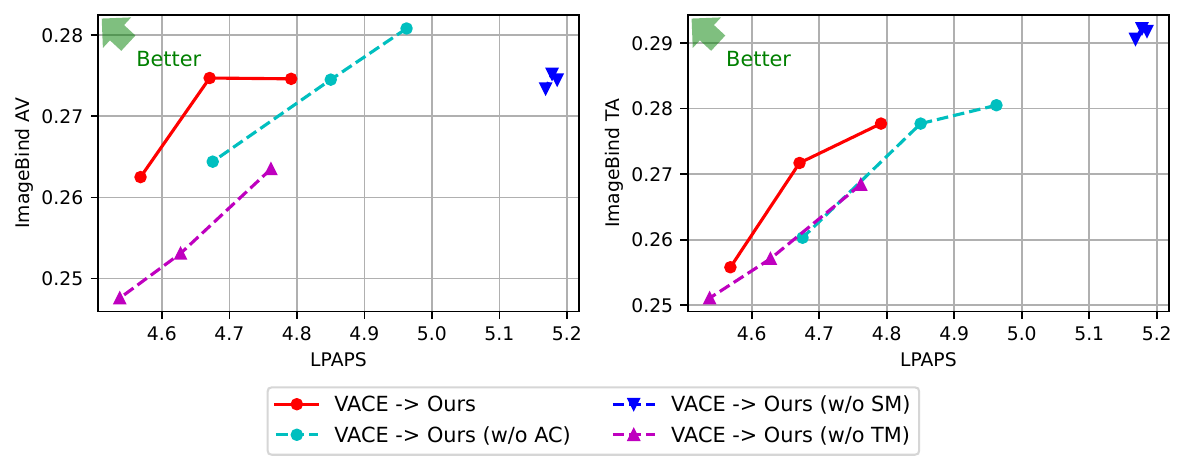}
    \caption{Ablation study. AC, SM, and TM denote adaptive conditioning, Synchformer feature modulation, and temporal masking, respectively.}
    \label{fig:ablation}
\end{figure}

We also performed an ablation study to clarify how much each newly introduced component contributes to enhancing the quality of edited audio. Specifically, we examine our method to remove the Synchformer feature modulation (SM) described in Section \ref{sec:architecture}, the temporal masking (TM) in Section \ref{sec:masking}, and the adaptive conditioning (AC) in Section \ref{sec:adaptive_conditioning}. In the case without AC, we used a fixed level of detail in all samples and varied it to control the overall degree of structure preservation.

Figure \ref{fig:ablation} shows the results of the ablation study. 
\begin{itemize}
    \item {\bf Without SM}, the generated sounds result in similar LPAPS across various levels of detail, while achieving high performance in both audio-visual alignment and text fidelity. This indicates that modulating Synchformer features is important for reflecting conditional acoustic information in the generated target audio.
    \item {\bf Without TM}, the quality of the generated sounds degrades in both audio-visual alignment and text fidelity. This highlights that augmenting the acoustic features using temporal masking contributes to effective training of our model, even though such masking is never used during inference.
    \item {\bf Without AC}, LPAPS is generally higher than with AC, while performance in audio-visual alignment and text fidelity remains similar. This shows that adaptively controlling the level of detail based on editability contributes to accurate preservation of the acoustic structure while maintaining the quality of the generated audio.
\end{itemize}

\paragraph{Limitations.} Since our pipeline is sequential, the generation quality can be degraded by errors in video editing. As shown in Fig. \ref{fig:main_results}, using more advanced video editing methods leads to better performance; thus, careful selection of the video editing method is important for our pipeline.

\section{Conclusion}

We introduced a novel audio-visual editing pipeline that first applies an existing video editing method, followed by audio editing that aligns with the video edits. To enable this, we developed a new video-to-audio model that uses the source audio as a conditional input. Our model adaptively extracts acoustic features from the source audio based on an editability score and incorporates them into the generation process to maximize retention of the original audio structure. Experimental results demonstrate that our model achieves high audio-visual alignment in the edited audio-video pairs while effectively preserving the acoustic characteristics of the source audio. 
As future work, we plan to extend the pipeline to support a broader range of editing tasks, such as object deletion and addition, which may require further advances in acoustic structure preservation.

\clearpage

%
%
\bibliographystyle{splncs04}
\bibliography{main}
\end{document}